\documentclass[preprint,12pt]{elsarticle}

\usepackage{amssymb}
\usepackage{amsmath}

\usepackage{algorithm}
\usepackage{algorithmicx}
\usepackage{algpseudocode}

\usepackage{xcolor}
\usepackage{amsmath}
\usepackage{amssymb}
\usepackage{multicol}
\usepackage{varwidth}

\usepackage{hyperref}
\usepackage{amsmath}
\renewcommand{\citep}[1]{(\citeyear{#1})}
\journal{Pattern Recognition}

\begin{document}

\begin{frontmatter}







\title{Self-Consistent Nested Diffusion Bridge for Accelerated MRI Reconstruction}
\author[1,7,8]{Tao Song\corref{eq}}
\ead{tsong22@m.fudan.edu.cn}
\author[2]{Yicheng Wu\corref{eq}}
\cortext[eq]{Equal Contribution}
\author[3]{Minhao Hu}
\author[4]{Xiangde Luo}
\author[6]{Fang Nie}
\author[5]{Guoting Luo}
\author[4]{Guotai Wang}
\author[1]{Yi Guo}
\author[1]{Feng Xu}
\author[7]{Shaoting Zhang}
\ead{Zhangshaoting@pjlab.org.cn}
        
\address[1]{School of Information Science and Technology, Fudan University, Shanghai, China}
\address[2]{Department of Data Science \& AI, Faculty of Information Technology, Monash University, Melbourne, Australia}
\address[3]{Nuffield Department of Clinical Neurosciences, University of Oxford, London, UK}
\address[4]{School of Mechanical and Electrical Engineering, University of Electronic Science and Technology of China, Chengdu, China}
\address[5]{Department of Radiology, Sichuan Provincial People's Hospital, Chengdu, China}
\address[6]{Department of Radiology, Zhongda Hospital, Medical School, Southeast University, Nanjing, China}
\address[7]{Shanghai AI Lab, Shanghai, China}
\address[8]{Sensetime Research, Shanghai, China}

\begin{abstract}

Accelerated MRI reconstruction plays a vital role in reducing scan time while preserving image quality. While most existing methods rely on complex-valued image-space or k-space data, these formats are often inaccessible in clinical practice due to proprietary reconstruction pipelines, leaving only magnitude images stored in DICOM files. To address this gap, we focus on the underexplored task of magnitude-image-based MRI reconstruction. Recent advancements in diffusion models, particularly denoising diffusion probabilistic models (DDPMs), have demonstrated strong capabilities in modeling image priors. However, their task-agnostic denoising nature limits performance in source-to-target image translation tasks, such as MRI reconstruction. In this work, we propose a novel Self-Consistent Nested Diffusion Bridge (SC-NDB) framework that models accelerated MRI reconstruction as a bi-directional image translation process between under-sampled and fully-sampled magnitude MRI images. SC-NDB introduces a nested diffusion architecture with a self-consistency constraint and reverse bridge diffusion pathways to improve intermediate prediction fidelity and better capture the explicit priors of source images. Furthermore, we incorporate a Contour Decomposition Embedding Module (CDEM) to inject structural and textural knowledge by leveraging Laplacian pyramids and directional filter banks. Extensive experiments on the fastMRI and IXI datasets demonstrate that our method achieves state-of-the-art performance compared to both magnitude-based and non-magnitude-based diffusion models, confirming the effectiveness and clinical relevance of SC-NDB.

\end{abstract}





\end{frontmatter}


\section{Introduction}
Magnetic resonance imaging (MRI) is widely used in clinical practice due to its ability to visualize various human regions with diverse contrasts. However, MRI scanning is inherently slow (\textit{e.g.,} a 3.0T brain examination typically takes 35–45 minutes \cite{sartoretti2019reduction}), resulting in high costs and increased susceptibility to motion artifacts \cite{zaitsev2015motion}. Consequently, accelerated MRI reconstruction has garnered significant attention, leading to the development of various deep learning-based approaches aimed at reducing scanning time and improving reconstruction quality \cite{deep2,jeelani2018image,fabian2021data,deep1,chen2021wavelet,peng2022towards,luo2023bayesian}.

\begin{figure}[tbp]
    \centering
    \includegraphics[width=0.8\textwidth]{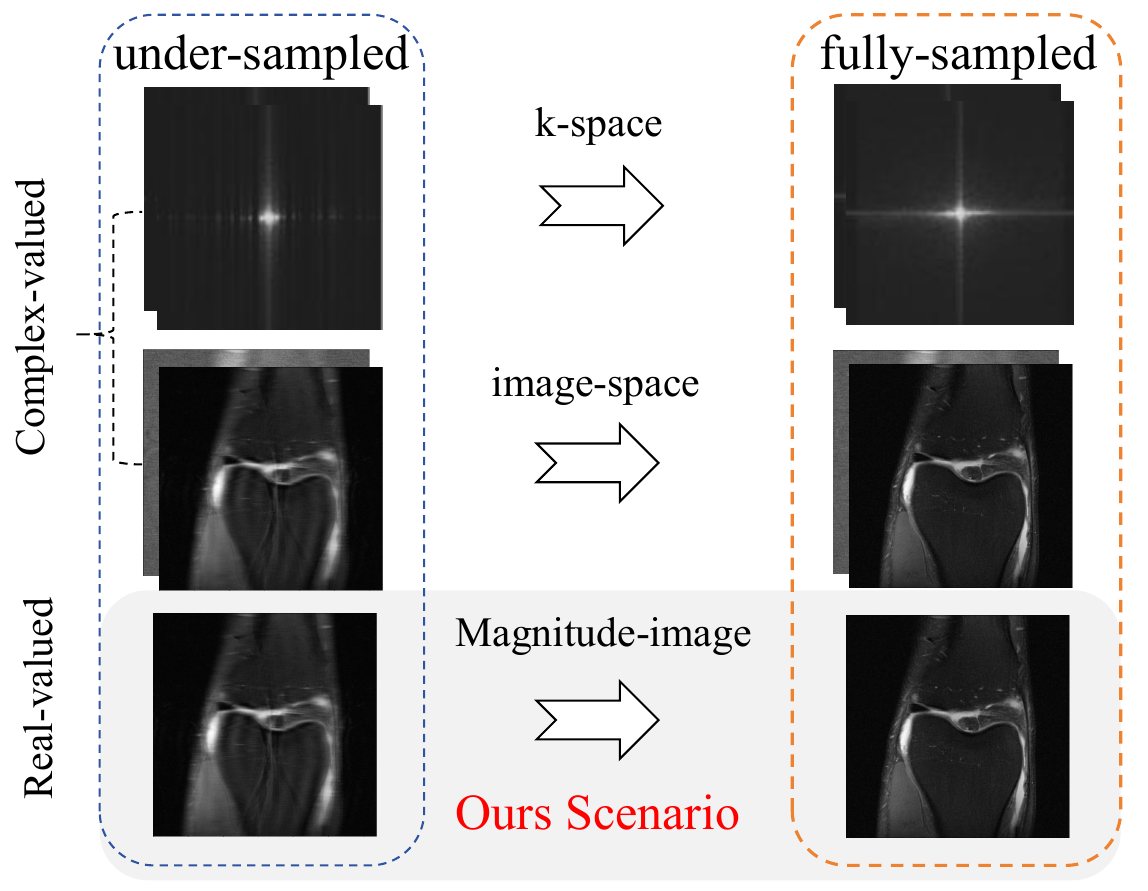}
    \caption{Categorized reconstruction methods based on data types: image-space, k-space, and magnitude-image. Our proposed method specifically focuses on the magnitude image scenario.}
    \label{fig:scenario}
\end{figure}

Accelerated MRI reconstruction methods can be categorized based on the type of input data into image-space, k-space, and magnitude-image approaches. The distinction between magnitude images and raw data is illustrated in Fig.~\ref{fig:scenario}. While image-space and k-space data can be mutually converted through Fourier transform, magnitude-image cannot be reversibly transformed back to either image-space or k-space data. 

Most accelerated MRI reconstruction methods rely on complex-valued data (k-space or image-space)~\cite{ADOBI,chen2021wavelet,zhang2018ista,gao2025self,directnips23,hou2023fast,grappa,gungor2023adaptive, xie2022measurement, high-frequency, hou2022idpcnn, zheng2019cascaded,cui2021equilibrated}, which is derived from intermediate processes during MRI scanning. However, obtaining such raw data is clinically impractical since manufacturer-specific reconstruction pipelines are inherently required before DICOM file storage, which only preserves magnitude images. In practical scenarios, deep learning-based MRI accelerated reconstruction using magnitude-image~\cite{image-based} is also widely adopted in clinical applications, offering better generalization and processing speed~\cite{yang2024impact}. Therefore, this study primarily focuses on the relatively underexplored area of magnitude-image-based accelerated MRI reconstruction.

Many studies has adopted denoising diffusion probabilistic models (DDPMs) to capture explicit image priors~\cite{DDPM,IDDPM,scorediff,ddim}, enhancing training stability and improving image fidelity. In DDPMs, the forward diffusion process gradually degrades the target image by repeatedly adding Gaussian noise until it becomes pure noise. Starting from a randomly sampled noise image, the reverse diffusion process then progressively removes the noise using a neural network to recover the target image, conditioned on the source image. However, DDPMs learn a task-agnostic denoising transformation from noise to the target image, which can weaken the guidance from the source image and thus may underperform in source-to-target image translation tasks.

Recently, the bridge diffusion model has emerged as a promising approach for enhancing source image relevance in image translation tasks by enabling direct transformations between two distinct distribution images~\cite{bbdm,I2SB,brigediffusion}. To this end, bridge diffusion defines the start and end points of the forward diffusion process based on the source and target images, respectively. Sampling begins directly from the source images, and the reverse process incrementally transforms it toward the target image. Unlike conventional bridge diffusion models, ~\citet{selfRDB} introduces a novel noise scheduling strategy in the forward process, where the variance monotonically increases according to the end state corresponding to the noise-added source images. This design aims to capture soft priors from the source modality and improve generalization.

While existing bridge diffusion methods focus primarily on innovations in noise scheduling, limited attention has been paid to self-consistency in the reverse diffusion process during training. To address this, we propose a novel Self-Consistent Nested Diffusion Bridge (SC-NDB) to enhance source-to-target image translation performance. In contrast to conventional bridge diffusion models, SC-NDB incorporates both a reverse bridge diffusion model and a bi-directional translation process during training, aiming to improve the accuracy of intermediate state predictions. This allows the model to more accurately capture explicit priors from the source image, thus facilitating more effective information transfer between source and target images.

The difference between fully-sampled and under-sampled MRI images is primarily manifested in the high-frequency components. To explicitly incorporate structural and textural knowledge specific to MRI, we introduce a Contour Decomposition Embedding Module (CDEM) to further boost performance. This module decomposes low-level features from each time-step image using iterative Laplacian pyramids~\cite{lp} and directional filter banks~\cite{DFB1,DFB2}, extracting structural and textural information and integrating it into the denoising network, thereby enhancing the network’s ability to preserve anatomical details.

Extensive validation was conducted on the publicly available fastMRI and IXI datasets. Our results clearly demonstrate the superior performance of SC-NDB compared to competitive magnitude-image-based diffusion models and non-magnitude-image-based models (k-space and image-space), confirming its effectiveness in MRI reconstruction tasks.

Overall, the contributions in this paper are three-fold:
\begin{itemize}
\item[1)] We propose a Self-Consistent Nested Diffusion Bridge (SC-NDB), which formulates accelerated MRI reconstruction as a bi-directional translation process between under-sampled and fully-sampled MRI magnitude images. This process incorporates a self-consistency loss via a nested diffusion bridge to more accurately capture explicit priors from the source images.

\item[2)] We propose a Contourlet Decomposition Embedding Module (CDEM) that leverages structural texture knowledge extracted from the frequency domain to mitigate the structural discrepancies caused by high-frequency misalignment between under-sampled and fully-sampled MR magnitude images.

\item[3)] Extensive experiments on the public fastMRI and IXI datasets demonstrate the effectiveness of our proposed Self-Consistent Diffusion Bridge Model for in-distribution and out-of-distribution accelerated MRI reconstruction.
\end{itemize}

\begin{figure*}[h!]
    \centering
    \includegraphics[width=1.0\textwidth]{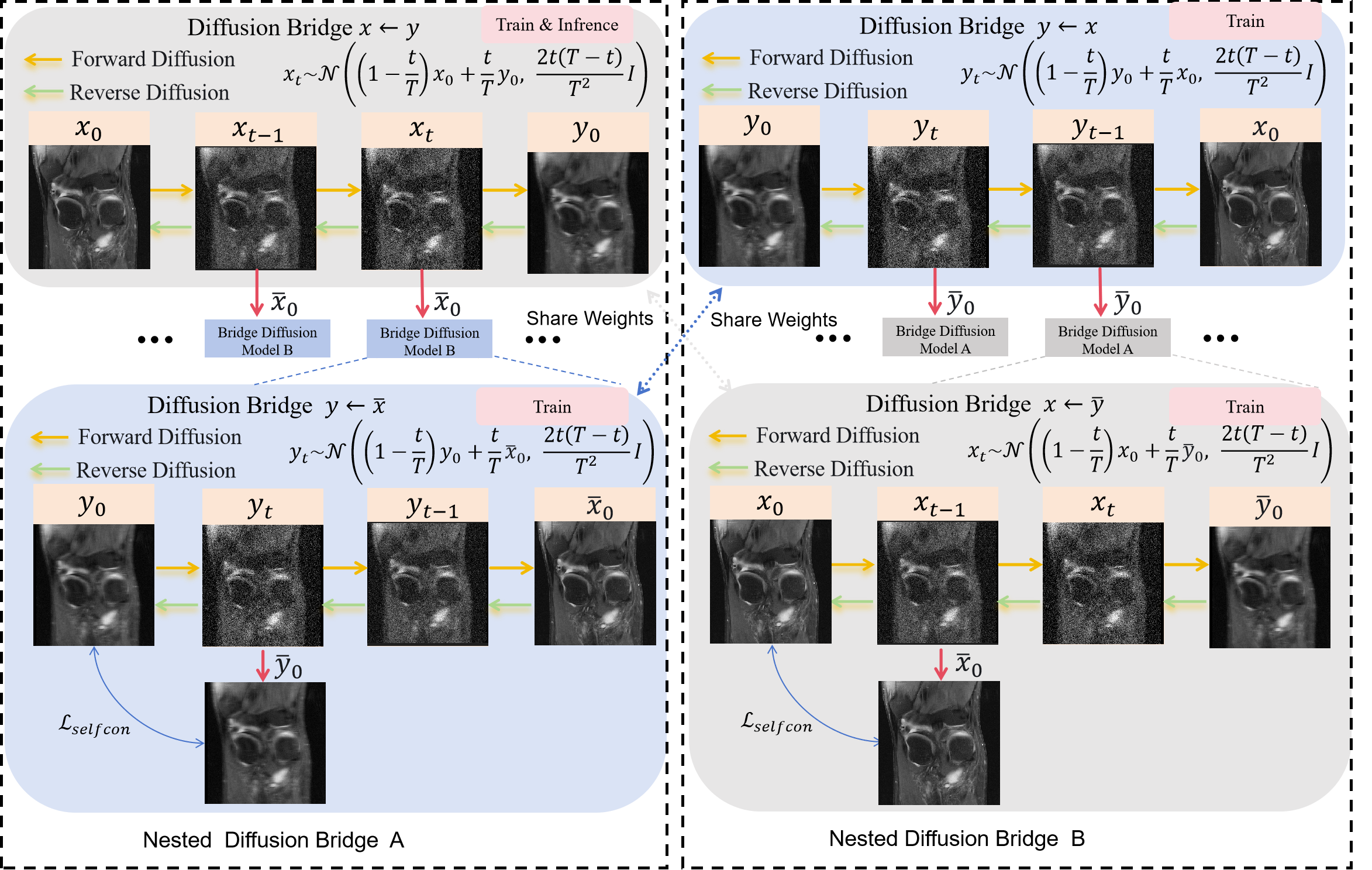}
    \caption{The framework of the SC-NDB involves two nested diffusion bridges between deterministically under-sampled and fully-sampled MR magnitude images.}
    \label{fig:framework}
\end{figure*}

\section{Related Work}
\subsection{Accelerated MRI Reconstruction}
For accelerated MRI reconstruction, traditional approaches primarily include Sensitivity Encoding (SENSE) \cite{sense} and GeneRalized Autocalibrating Partially Parallel Acquisitions (GRAPPA) \cite{grappa}. SENSE leverages the spatial sensitivity of multiple receiver coils to achieve faster imaging by reducing the acquisition time. On the other hand, GRAPPA reconstructs under-sampled data by estimating missing k-space data using kernel-based interpolation, enhancing the final image quality and reducing artifacts associated with parallel imaging techniques.
\begin{figure}[tbp]
    \centering
    \includegraphics[width=0.8\textwidth]{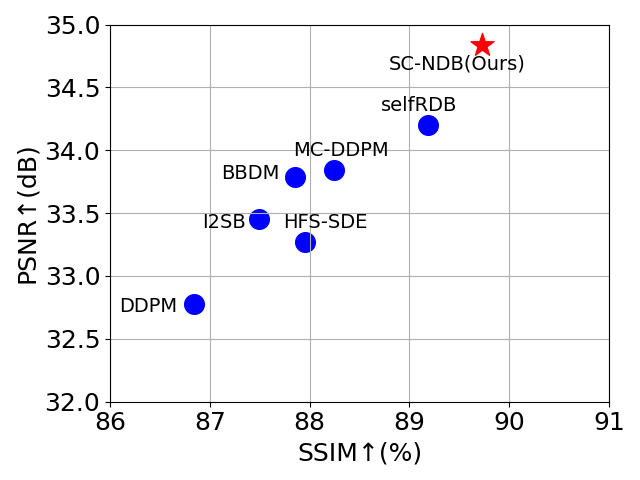}
    \caption{Comparisons with state-of-the-art methods in terms of PSNR and SSIM results on the fastMRI knee dataset.}
    \label{fig:sota_scatter}
\end{figure}

With the success of data-driven approaches, deep learning has been widely utilized in accelerated MRI reconstruction. \citet{lv2021pic,deep1} integrate deep learning with parallel imaging algorithms. \cite{deep1} combined deep learning with the GRAPPA reconstruction algorithm. \cite{ista-net} transform the Iterative Shrinkage-Thresholding Algorithm (ISTA) \cite{ista} into a deep neural network to optimize compressive sensing (CS) for accelerated MRI reconstruction. \cite{shitrit2017accelerated} utilized an adversarial neural network to generate the missing k-space data. The generated and missing k-space data are subsequently transformed via Fourier transform and then fed into the discriminator for evaluation. \cite{cole2020unsupervised} only requires under-sampled k-space data. The generator produces under-sampled k-space data from the generated image and the discriminator evaluates the k-space measurements instead of the MR image and provides the learned gradients back to the generator.
Due to the high-fidelity generation characteristics of diffusion models, they have been widely applied in accelerated MRI reconstruction in recent years. \cite{peng2022towards} introduced an unconditional diffusion model, which is trained to generate coil-combined MR image samples derived from fully-sampled data. \cite{gungor2023adaptive} proposed a conditional diffusion model for accelerated MRI reconstruction based on adaptive diffusion priors. To fully utilize the invariant nature of the low-frequency region in k-space during accelerated sampling, \cite{high-frequency} introduced a conditional diffusion model based on high-frequency diffusion. Similarly, \cite{xie2022measurement} proposed a measurement-conditioned diffusion model, which avoids diffusion in the under-sampled mask region rather than only outside the low-frequency area, thereby preserving more prior information. To further enhance the guidance derived from the source images, bridge diffusion models have recently been introduced into the task of MRI accelerated reconstruction. \citet{ADOBI} proposed a bridge diffusion model that adaptively calibrates an unknown forward model to reinforce measurement consistency throughout the sampling iterations, addressing the blind inverse problem in MRI accelerated reconstruction. \citet{gao2025self} introduced a self-supervised bridge diffusion approach that directly trains on available noisy measurements without the need for any high-quality reference images.

Although diffusion models for accelerated MRI reconstruction have evolved from conditional diffusion to bridge diffusion conditioned on source images, they are typically built upon complex-valued image-space or k-space data as source and target domains. In contrast, diffusion modeling based on magnitude images (widely stored in clinical DICOM formats) has been rarely explored.

\subsection{Diffusion Model}
Recent advancements in diffusion models, such as those by \cite{DDPM, IDDPM, scorediff,colddiff}, have significantly enhanced image generation, surpassing the performance of GANs \cite{goodfellow2014generative}. These successes are attributed to key design choices, including network architecture \cite{scorediff, karras2022}, optimized noise schedules \cite{IDDPM, karras2022}, improved sampling techniques \cite{ddim, lu2022fast, zhang2022fast}, and advanced guidance methods \cite{IDDPM,ho2022classifier}.

Bridge diffusion models \cite{bbdm,brigediffusion,I2SB,selfRDB} have been proposed for image translation tasks, where the diffusion process is constructed between paired images. This approach starts reverse diffusion from a deterministic image (without noise), making the sampling process more stable. Existing studies on diffusion bridges have primarily focused on innovations in noise scheduling, while overlooking the self-consistency of intermediate states. This limitation hinders the accurate capture of prior distributions from the source images. To address this, we propose a Self-Consistent Nested Diffusion Bridge (SC-NDB) to model the translation from under-sampled MRI images to fully-sampled MRI images.

\section{Methodology}
Fig.~\ref{fig:framework} provides an overview of our Self-Consistent Nested Diffusion Bridge, where two nested diffusion bridges, A and B, are constructed using four distinct diffusion bridges. Specifically, the diffusion bridges $\boldsymbol{x} \leftarrow \boldsymbol{y}$ and $\boldsymbol{x} \leftarrow \boldsymbol{\overline{y}}$ share parameters $\theta_1$, while $\boldsymbol{y} \leftarrow \boldsymbol{x}$ and $\boldsymbol{y} \leftarrow \boldsymbol{\overline{x}}$ share parameters $\theta_2$. Here, $\boldsymbol{x}_0$ and $\boldsymbol{y}_0$ denote the fully-sampled and under-sampled magnitude images. $\to$ and $\gets$ denote the forward and reverse processes in each diffusion model, respectively.

\subsection{Bridge Diffusion Model}
Given the inconsistent predictions of typical diffusion models, we exploit the Brownian bridge diffusion model \cite{bbdm} for the deterministic image-to-image transformation. The state distribution at each time step $t$ of a forward process (\textit(e.g.,) $\boldsymbol{x} \to \boldsymbol{y}$) can be expressed as:
\begin{equation}
    q(\boldsymbol{x}_t|\boldsymbol{x}_0,\boldsymbol{y}_0)=\boldsymbol{N}(\boldsymbol{x}_t;(1-m_{t})\boldsymbol{x}_0+m_{t}\boldsymbol{y}_0,\sigma_t\boldsymbol{I})
\end{equation}
\begin{equation}
        m_{t}=\frac{t}T,  \sigma_t=2(m_{t}-(m_{t})^2)\nonumber
\end{equation}
where $\boldsymbol{x}_0\sim\boldsymbol{q}_{data}(\boldsymbol{x}_0)$, and $\boldsymbol{y}_0\sim\boldsymbol{q}_{data}(\boldsymbol{y}_0)$. $T$ is the total number of iteration steps and $\sigma_t$ denotes the variance. Here, the variances at the starting and ending states, \textit(i.e.,) $\sigma_0$ and $\sigma_T$ are 0, avoiding the random predictions. Meanwhile, $\sigma_{T/2}=0.5$ is bounded, which could smooth the model training \cite{bbdm}.

The intermediate state $\boldsymbol{x}_t$ in the forward process, transitioning from the starting state $\boldsymbol{x}_0$ to the ending state $\boldsymbol{y}_0$, can be represented as 
\begin{equation}
    \boldsymbol{x}_t=(1-m_{t})\boldsymbol{x}_0+m_{t}\boldsymbol{y}_0+\sqrt{\sigma_t}\boldsymbol{\epsilon}\label{con:farword}
\end{equation}
where $\boldsymbol{\epsilon}\sim\mathcal{N}(\mathbf{0},\boldsymbol{I})$.

Then, for the reverse process, the Brownian Bridge model initiates directly from image $\boldsymbol{y}_0$. Following the principle of denoising diffusion methods, the reverse process ($\boldsymbol{x} \gets \boldsymbol{y}$) aims to predict $\boldsymbol{x}_{t-1}$ based on $\boldsymbol{x}_{t}$:
\begin{equation}
\begin{aligned}
&p_\theta(\boldsymbol{x}_{t-1}|\boldsymbol{x}_t,\boldsymbol{y}_0)=\mathcal{N}(\boldsymbol{x}_{t-1};\boldsymbol{\mu}_\theta(\boldsymbol{x}_t,t),\tilde{\sigma}_t\boldsymbol{I})
\\ 
&s.t., \    \tilde{\sigma}_t= \sigma_{t|t-1}\frac{\sigma_{t-1}}{\sigma_{t}} \\
 & \sigma_{t|t-1}=\sigma_t-\sigma_{t-1}\frac{(1-m_{t})^2}{(1-m_{t-1})^2}
\end{aligned}
\end{equation}
where $\boldsymbol{\mu}_\theta(\boldsymbol{x}_t,t)$  represents the predicted mean value of noise at the $t$-th steps, and $\tilde{\sigma}_t$ denotes noise variances in the reverse process. 

Finally, the loss function is formulated by optimizing the Evidence Lower Bound (ELBO) for the Brownian Bridge diffusion process:
\begin{equation}
\mathbb{E}_{\boldsymbol{x}_0,\boldsymbol{y}_0,\boldsymbol{\epsilon}}[||(\underbrace{m_t(\boldsymbol{y}_0-\boldsymbol{x}_0)+\sqrt{\sigma_t}\boldsymbol{\epsilon})}_{\text{objective}}-\boldsymbol{\epsilon}_\theta(\boldsymbol{x}_t,t)||]\label{con:minimize}
\end{equation}

Therefore, let $\boldsymbol{\theta1}$ and $\boldsymbol{\theta2}$ represent the denoising networks for $\boldsymbol{x}\gets \boldsymbol{y}$ and $\boldsymbol{y}\gets \boldsymbol{x}$ processes, respectively. The diffusion losses $\mathcal{L}_{\boldsymbol{x}\gets \boldsymbol{y}}^{rec}$ and $\mathcal{L}_{\boldsymbol{y}\gets \boldsymbol{x}}^{rec}$ at step $t$ are:

\begin{equation}
\begin{aligned}
& \mathcal{L}_{\boldsymbol{x}\gets \boldsymbol{y}}^{rec} = \mathbb{E}_{\boldsymbol{x}_{0},\boldsymbol{y}_{0},\boldsymbol{\epsilon}}[||(m_{t}(\boldsymbol{y}_{0}-\boldsymbol{x}_{0})+\sqrt{\sigma_t}\boldsymbol{\epsilon})-\boldsymbol{\epsilon}_{\theta1}(\boldsymbol{x}_t,t)||] \\
& \mathcal{L}_{\boldsymbol{y}\gets \boldsymbol{x}}^{rec} = \mathbb{E}_{\boldsymbol{y}_{0},\boldsymbol{x}_{0},\boldsymbol{\epsilon}}[||(m_{t}(\boldsymbol{x}_{0}-\boldsymbol{y}_{0})+\sqrt{\sigma}_t\boldsymbol{\epsilon})-\boldsymbol{\epsilon}_{\theta2}(\boldsymbol{y}_t,t)||].
\end{aligned}
\end{equation}
Note that, Equation \eqref{con:minimize} theoretically indicates that with sufficiently low loss, the original target image $\boldsymbol{x}_0$ can be estimated at any time $t$ given a noisy image $\boldsymbol{x}_t$. Therefore, the reconstructed $\boldsymbol{\bar{x}}_0$ based on $\boldsymbol{y}_0$ at $t$ step can be calculated using the generation function only in training process:

\begin{equation}
    \boldsymbol{\bar{x}}_0 = \boldsymbol{x}_t - \epsilon_\theta(\boldsymbol{x}_t,t)\label{con:predx0}
\end{equation}
Note, the $\boldsymbol{\bar{x}}_0$ obtained during inference is generated through inverse diffusion sampling, following the same sampling procedure as described in \cite{bbdm}.

\subsection{Self-Consistent Nested Diffusion Bridge}
To more accurately capture the prior distribution of source images, a new diffusion bridge is nested to enforce self-consistency. Specifically, Equation \eqref{con:predx0} indicates that, during training, the target image $\boldsymbol{x}_0$ or $\boldsymbol{y}_0$ can be estimated by a noisy state $\boldsymbol{x}_{t1}$ and $\boldsymbol{y}_{t1}$, respectively. Therefore, the reconstructed fully-sampled and under-sampled MR magnitude images $\boldsymbol{\bar{x}}_0$ and $\boldsymbol{\bar{y}}_0$ become:
\begin{equation}
\begin{aligned}
    \boldsymbol{\bar{x}}_0 = \boldsymbol{x}_{t1} - \epsilon_{\theta1}(\boldsymbol{x}_{t1},t1) \\
    \boldsymbol{\bar{y}}_0 = \boldsymbol{x}_{t1} - \epsilon_{\theta2}(\boldsymbol{y}_{t1},t1).
\end{aligned}
\end{equation}

We then designate the reconstructed samples as the new end states of the new diffusion bridges $\boldsymbol{y} \leftarrow \boldsymbol{\overline{x}}$ and $\boldsymbol{x} \leftarrow \boldsymbol{\overline{y}}$, respectively. Consequently, the intermediate states $\boldsymbol{x}'_{t2}$ and $\boldsymbol{y}'_{t2}$ are defined as follows:

\begin{equation}
\begin{aligned}
&\boldsymbol{y}'_{t2}=(1-m_{t2})\boldsymbol{y}_0+m_{t2}\boldsymbol{\bar{x}}_0+\sqrt{\sigma_{t2}}\boldsymbol{\epsilon} \\
&\boldsymbol{x}'_{t2}=(1-m_{t2})\boldsymbol{x}_0+m_{t2}\boldsymbol{\bar{y}}_0+\sqrt{\sigma_{t2}}\boldsymbol{\epsilon}\label{con:farword2}.
\end{aligned}
\end{equation}

Similarly, the reverse processes of the new diffusion bridges $\boldsymbol{y} \gets \boldsymbol{\bar{x}}$ and $\boldsymbol{x} \gets \boldsymbol{\bar{y}}$ result in the newly reconstructed images $\boldsymbol{\bar{\bar{y}}}_0$ and $\boldsymbol{\bar{\bar{x}}}_0$, respectively. These are given by:

\begin{equation}
\begin{aligned}     
&\boldsymbol{\bar{\bar{y}}}_0 = \boldsymbol{y}'_{t2} - \epsilon_{\theta2}(\boldsymbol{y}'_{t2},t2)
\\
&\boldsymbol{\bar{\bar{x}}}_0 = \boldsymbol{x}'_{t2} - \epsilon_{\theta1}(\boldsymbol{x}'_{t2},t2).
\end{aligned}
\end{equation}

Therefore, a self-consistent loss can be used to reduce the distance between $\boldsymbol{x}_{0}$ and $\boldsymbol{\bar{\bar{x}}}_0$, and between $\boldsymbol{y}_{0}$ and $\boldsymbol{\bar{\bar{y}}}_0$, respectively as:



\begin{equation}
\begin{aligned}
    \mathcal{L}_{\boldsymbol{x}}^{selfcon}(\theta1, \theta2)=\mathbb{E}_{\boldsymbol{x}_0,\boldsymbol{y}_0}[||\boldsymbol{x}_0-\boldsymbol{\bar{\bar{x}}}_0||]
\\
   \mathcal{L}_{\boldsymbol{y}}^{selfcon}(\theta1, \theta2)=\mathbb{E}_{\boldsymbol{x}_0,\boldsymbol{y}_0}[||\boldsymbol{y}_0-\boldsymbol{\bar{\bar{y}}}_0||].
\end{aligned}
\end{equation}
\begin{algorithm}
\caption{Pseudo training codes of our Self-Consistent Nested Diffusion Bridge for accelerated MRI reconstruction}
\begin{algorithmic}[1]
\Repeat
  \State Paired data $\boldsymbol{x}_{0}\sim\boldsymbol{q}(\boldsymbol{x}_0)$, $\boldsymbol{y}_{0}\sim\boldsymbol{q}(\boldsymbol{y}_0)$
  \State Timestep $t1\sim Uniform(1,...,T)$
  \State Timestep $t2\sim Uniform(1,...,T)$
  \State Gaussian noise $\epsilon\sim\mathcal{N}(\mathbf{0},\mathbf{I})$
  \State Forward diffusion $\boldsymbol{x}_0\to \boldsymbol{y}_0$ \Statex \hspace{1cm}$\boldsymbol{x}_{t1}=(1-m_{t1})\boldsymbol{x}_{0}+m_{t1}\boldsymbol{y}_{0}+\sqrt{\sigma_{t1}}\boldsymbol{\epsilon} $
  \State Computing $ \boldsymbol{\bar{x}}_0 = \boldsymbol{x}_{t1} - \epsilon_{\theta1}(\boldsymbol{x}_{t1},t1)$
  \State Forward diffusion $\boldsymbol{y}_0\to \boldsymbol{x}_0$ \Statex
  \hspace{1cm}$\boldsymbol{y}_{t1}=(1-m_{t1})\boldsymbol{y}_{0}+m_{t1}\boldsymbol{x}_{0}+\sqrt{\sigma_{t1}}\boldsymbol{\epsilon} $
 \State Computing $\boldsymbol{\bar{y}}_0 = \boldsymbol{y}_{t1} - \epsilon_{\theta2}(\boldsymbol{y}_{t1},t1)$
 
  \State New Forward diffusion $\boldsymbol{y}_0\to \boldsymbol{\bar{x}_0}$ \Statex \hspace{1cm}$ \boldsymbol{y}'_{t2}=(1-m_{t2})\boldsymbol{y}_0+m_{t2}\boldsymbol{\bar{x}}_0+\sqrt{\sigma_{t2}}\boldsymbol{\epsilon}$
   \State Computing $ \boldsymbol{\bar{\bar{y}}}_0 = \boldsymbol{y}'_{t2} - \epsilon_{\theta2}(\boldsymbol{y}'_{t2},t2)$
   \State New Forward diffusion $\boldsymbol{x}_0\to \boldsymbol{\bar{y}_0}$ \Statex \hspace{1cm}$ \boldsymbol{x}'_{t2}=(1-m_{t2})\boldsymbol{x}_0+m_{t2}\boldsymbol{\bar{y}}_0+\sqrt{\sigma_{t2}}\boldsymbol{\epsilon}$
   \State Computing $ \boldsymbol{\bar{\bar{x}}}_0 = \boldsymbol{x}'_{t2} - \epsilon_{\theta1}(\boldsymbol{x}'_{t2},t2)$
   
  \State \begin{varwidth}{\linewidth}Take gradient descent step on
      \Statex \hspace{1cm}$\triangledown_{\theta1}||(m_{t1}(\boldsymbol{y}_{0}-\boldsymbol{x}_{0})+\sqrt{\sigma_{t1}}\boldsymbol{\epsilon})-\boldsymbol{\epsilon}_{\theta1}(\boldsymbol{x}_{t1},t1)||$
      \Statex \hspace{1cm}$\triangledown_{\theta2}||(m_{t1}(\boldsymbol{x}_{0}-\boldsymbol{y}_{0})+\sqrt{\sigma_{t1}}\boldsymbol{\epsilon})-\boldsymbol{\epsilon}_{\theta2}(\boldsymbol{y}_{t1},t1)||$
       \Statex \hspace{1cm}$\triangledown_{\theta1, \theta2}||\boldsymbol{y}_0-\boldsymbol{\bar{\bar{y}}}_0||$
        \Statex \hspace{1cm}$\triangledown_{\theta1, \theta2}||\boldsymbol{x}_0-\boldsymbol{\bar{\bar{x}}}_0||$
    \end{varwidth}
\Until
\end{algorithmic}
\end{algorithm}

\begin{figure}[tbp]
    \centering
    \includegraphics[width=0.8\textwidth]{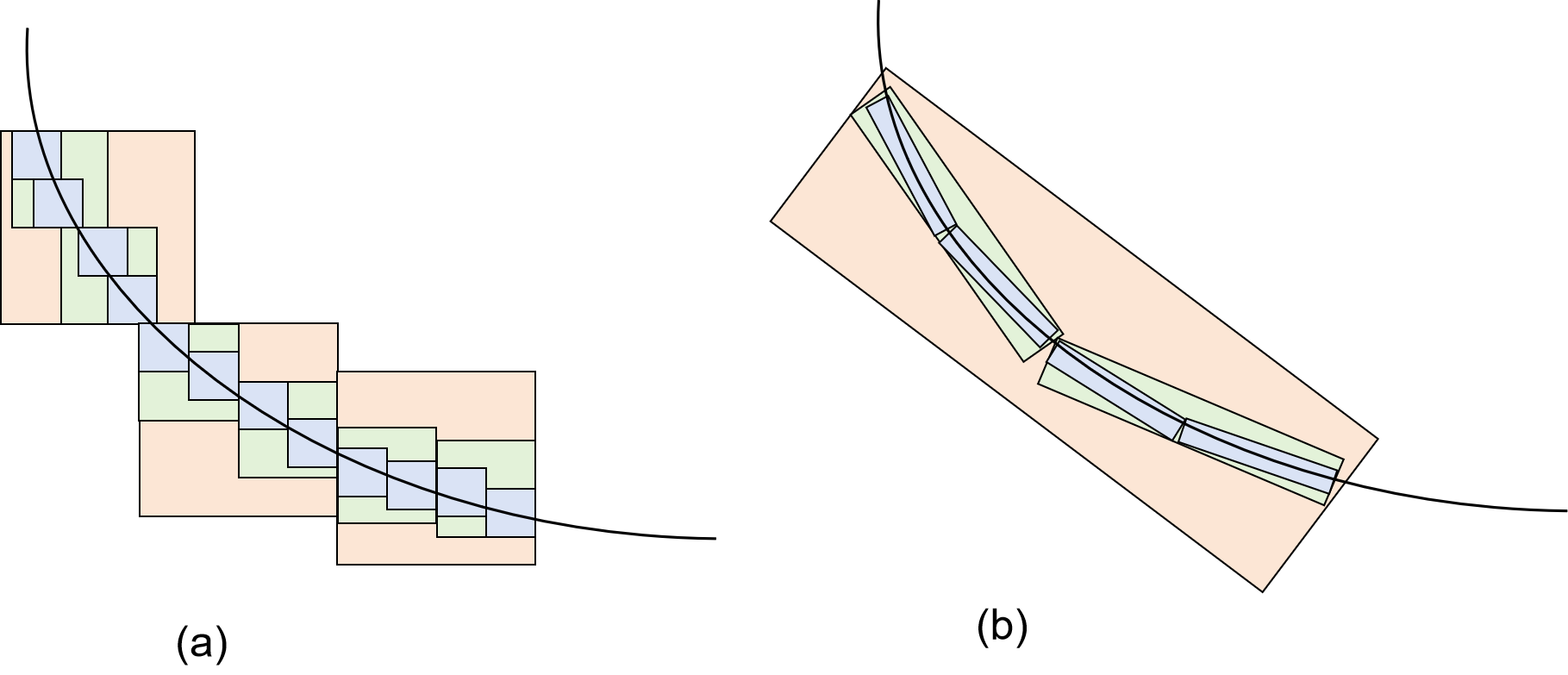}
    \caption{Visual comparison of sparsity representations using (a) traditional Wavelet Transform and (b) Contourlet Transform.}
    \label{fig:wave}
\end{figure}
Overall, the total loss is a weighted sum of reconstruction loss and self-consistent loss as
\begin{equation}
    \mathcal{L}_{total} = \lambda(\mathcal{L}_{\boldsymbol{x}}^{selfcon}+\mathcal{L}_{\boldsymbol{y}}^{selfcon})+( \mathcal{L}_{\boldsymbol{x}\gets \boldsymbol{y}}^{rec}+ \mathcal{L}_{\boldsymbol{y}\gets \boldsymbol{x}}^{rec})
\end{equation}
where $\lambda$ is a constant to balance losses for the model training. Algorithm 1 shows pseudo codes of our algorithm. Note that, only a denoising network $\boldsymbol{\theta1}$ is used for the model inference. Here, the model does not introduce additional inference time.

\subsection{Contourlet Decomposition Embedding Module}
As a signal representation tool, contourlets (analogous to edges in images) can effectively exploit geometric regularities to achieve sparse representations. Previous studies have utilized Fast Fourier Transform (FFT) or wavelet decomposition to capture texture information. However, the effectiveness of these methods is often limited in practice due to the lack of more precise design. As illustrated in Fig.~\ref{fig:wave}(a), wavelet decomposition produces a large number of inefficient point-based decompositions. In contrast, Fig.~\ref{fig:wave}(b) demonstrates that contourlet decomposition is capable of representing smooth contours using fewer coefficients by approximating them with linear segments. Consequently, contourlet-based representations are sparser and offer the advantages of directionality, locality, and bandpass characteristics.
\begin{figure}[tbp]
    \centering
    \includegraphics[width=0.9\textwidth]{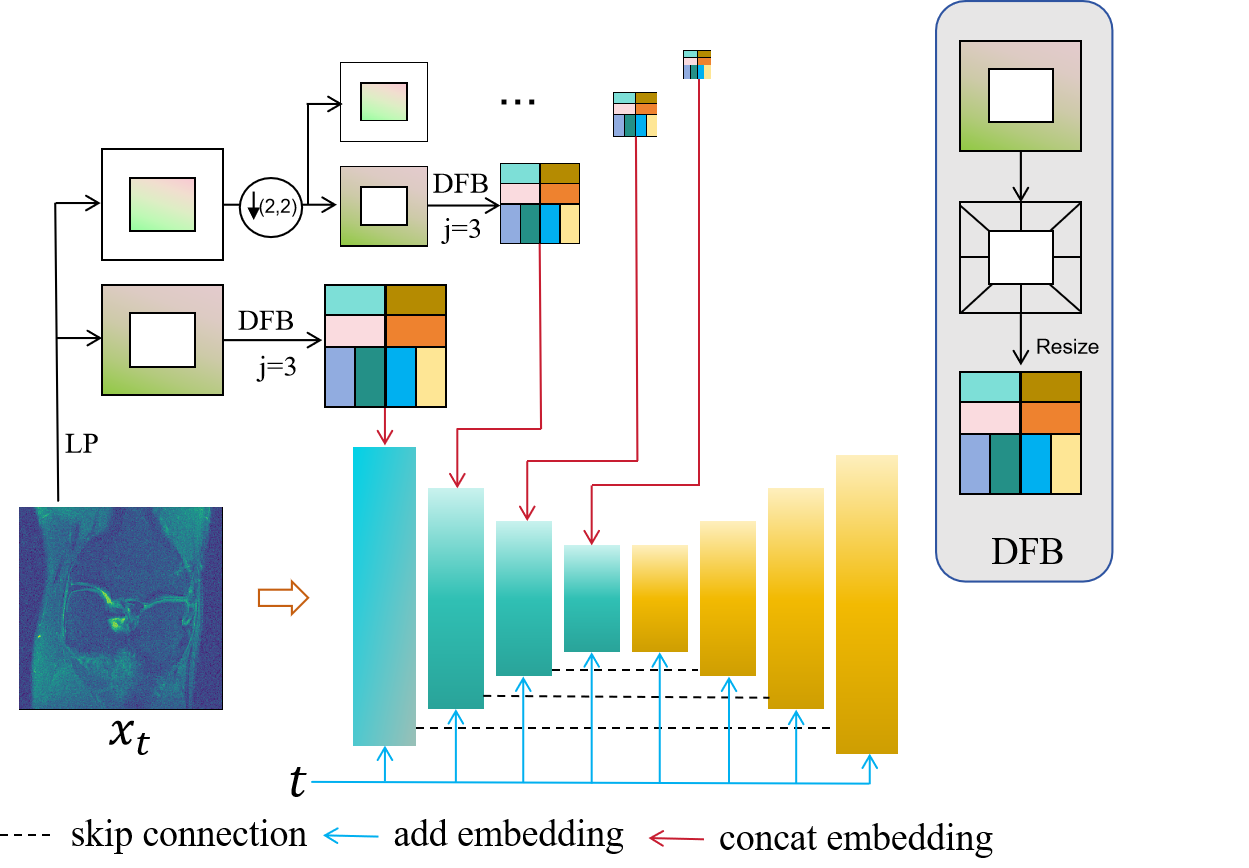}
    \caption{Denosing network with time and contourlet decomposition embedding.}
    \label{fig:contourlet}
\end{figure}

\begin{figure}[tbp]
    \centering
    \includegraphics[width=0.8\textwidth]{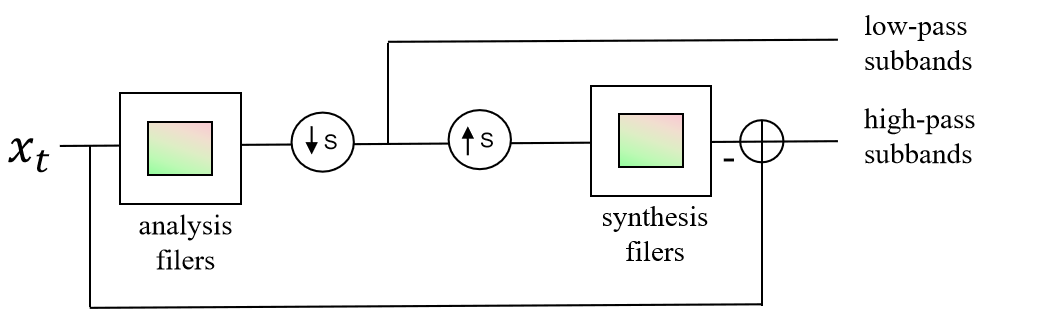}
    \caption{Details of the Laplacian Pyramid (LP) decomposition.}
    \label{fig:LP}
\end{figure}

The denoising networks of our SC-NDB model adopt a U-Net \cite{unet} with a time embedding. As illustrated in Fig.~\ref{fig:contourlet}, we further propose a contourlet decomposition embedding module to enhance the backbone. Generally, the blurring and artifacts in under-sampled MR magnitude images are caused by the missing high-frequency information in the frequency space. Contourlet decomposition employs the Laplacian Pyramid (LP)~\cite{lp} and Directional Filter Banks (DFB)~\cite{DFB1, DFB2} to process the low-pass images iteratively. The LP module performs a multi-scale decomposition to separate low- and high-frequency components, as illustrated in Fig.~\ref{fig:LP}. Specifically, the low-pass subband is generated by applying a low-pass analysis filter to the input $x_t$, followed by downsampling via a sampling matrix $\mathbf{S}$. The high-pass subband is then derived as the residual between the original input $x_t$ and the reconstructed low-pass subband, which is obtained through upsampling and a low-pass synthesis filter. To further enhance the representation of directional features, DFB is employed to decompose the high-pass subband into multiple directional components. By utilizing a $j$-level binary tree structure, DFB produces $2^j$ directional subbands, yielding compact and directionally sensitive representations in the 2D frequency domain.

\begin{figure}[tbp]
    \centering
    \includegraphics[width=0.8\textwidth]{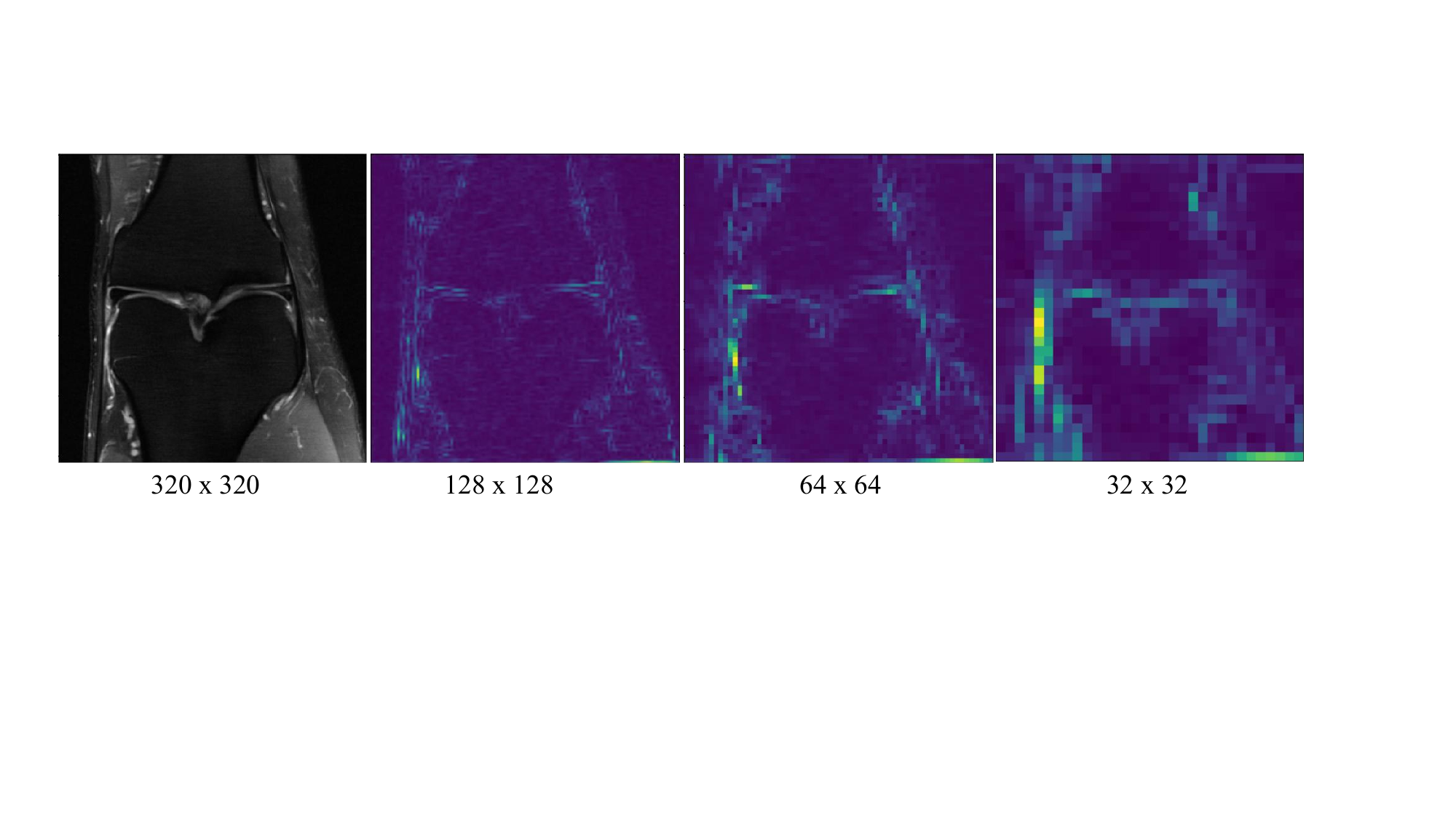}
    \caption{Contourlet decomposition features visualization with different levels.}
    \label{fig:contour_vis}
\end{figure}

For example, the frequency domain
is divided into $2^{3}$ directional sub-bands when $j=3$,
and sub-bands $0-3$ and $4-7$ correspond to the vertical and horizontal details, respectively.
\begin{equation}
    \begin{aligned}
F_{l},F_{h} & =LP(F_{l})\downarrow d \\
F_{bds} & =DFB(F_{h})\; n\in[1,j],
\end{aligned}
\end{equation}
where symbol $\downarrow$ is the downsampling operator and $d$ denotes the interlaced downsampling factor. The subscripts $l$ and $h$ represent the low-pass and high-pass components respectively, $bds$ indicates the bandpass directional subbands. 
 Fig~\ref{fig:contour_vis} illustrates the visualization of sub-band features at different levels obtained from the contourlet decomposition of $x_t$. It can be observed that contourlet decomposition effectively extracts high-frequency features of image edges. Then, the Contourlet Decomposition Embedding Module further resizes the features of multiple sub-bands at each level to match the feature size at the corresponding level, stacks them along the channel dimension, and then uses three convolution layers to align the channel number as corresponding level features.




\section{Experiments}
\subsection{Data and Implementation Details}
The experiments were conducted on the public benchmark fastMRI \footnote{\url{http://fastmri.med.nyu.edu}} and IXI datasets \footnote{\url{https://brain-development.org/ixi-dataset/}}. On the fastMRI dataset, we selected 360 individuals from the multi-coil knee dataset for training, 8 for validation, and 20 for testing. Following \cite{high-frequency}, the first six slices of each individual were excluded due to poor image quality. Consequently, the training, validation, and test sets contained 10,737, 244, and 576 2D images, respectively. Additionally, 218 images from 20 individuals from the fastMRI brain dataset were selected as an out-of-distribution set of images. All images were cropped to $320\times320$ and normalized to the range [0, 1]. From the IXI dataset, we selected 577 patients with T1 images, randomly split into training (500 patients, 44,935 2D images), validation (37 patients, 3,330 images), and test (40 patients, 3,600 images) sets. All images were cropped to $256\times256$ and normalized to the range [0, 1]. To simulate k-space undersampling, equally spaced Cartesian undersampling in the phase encoding direction was applied, with acceleration factors set to 4 and 8. The final MR magnitude images, both under-sampled and fully sampled, were generated from the corresponding k-space data through inverse Fourier transform and root sum of squares operations.
\begin{figure*}[h!]
    \centering
    \includegraphics[width=1.0\textwidth]{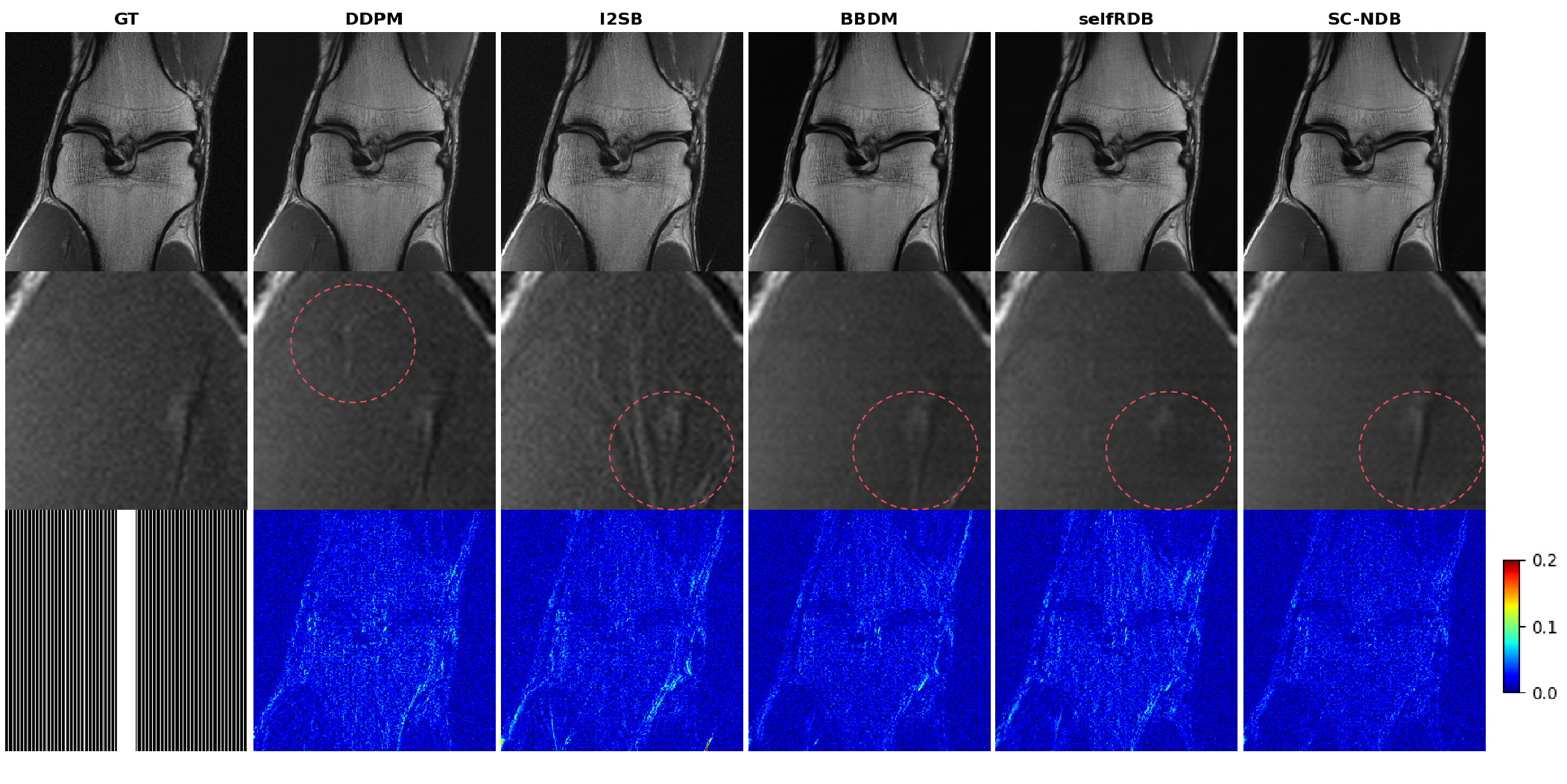}
    \caption{The reconstruction results of magnitude images generated from multi-coil knee fast MRI with a 4× acceleration factor under uniform Cartesian undersampling are presented (in-distribution). The first row displays the ground truth and the reconstructions from DDPM, I2SB, BBDM, selfRDB and SC-NDB (Ours). The second row presents the enlarged view of the ROI, and the third row shows the error map of the reconstruction. The lower left corner contains the undersampling mask.}
    \label{fig:knee_vis}
\end{figure*}
For training, the total number of iterations is 200k, using the AdamW optimizer with a learning rate of 1e-4 and a batch size of 28. For other diffusion methods (excluding AdaDiff), the training and sampling steps are set to 1000 and 200, respectively. In AdaDiff, the first stage employs an unconditional pretraining diffusion model with 1000 time steps, while the second stage uses 200 steps for iterative solving based on data consistency \cite{zheng2019cascaded}. For our SC-NDB, both the training and sampling steps are set to 20, with the loss weight $\lambda$ set to 1. All experiments were conducted in the same environment using 4 NVIDIA Tesla V100 GPUs.

The comparison was conducted by the proposed SC-NDB model and  AdaDiff \cite{gungor2023adaptive}, MC-DDPM \cite{xie2022measurement}, HFS-SDE \cite{high-frequency}, DDPM \cite{DDPM}, BBDM \cite{bbdm}, I2SB \cite{I2SB}, and selfRDB \cite{selfRDB}, where AdaDiff, MC-DDPM, and HFS-SDE are non-magnitude-image-based approaches. We further use the widely used Peak Signal-to-Noise Ratio (PSNR), Structural Similarity Index (SSIM) and Normalized Mean Square Error (NMSE) for the measurement of reconstruction quality.

\begin{figure*}[h!]
    \centering
    \includegraphics[width=1.0\textwidth]{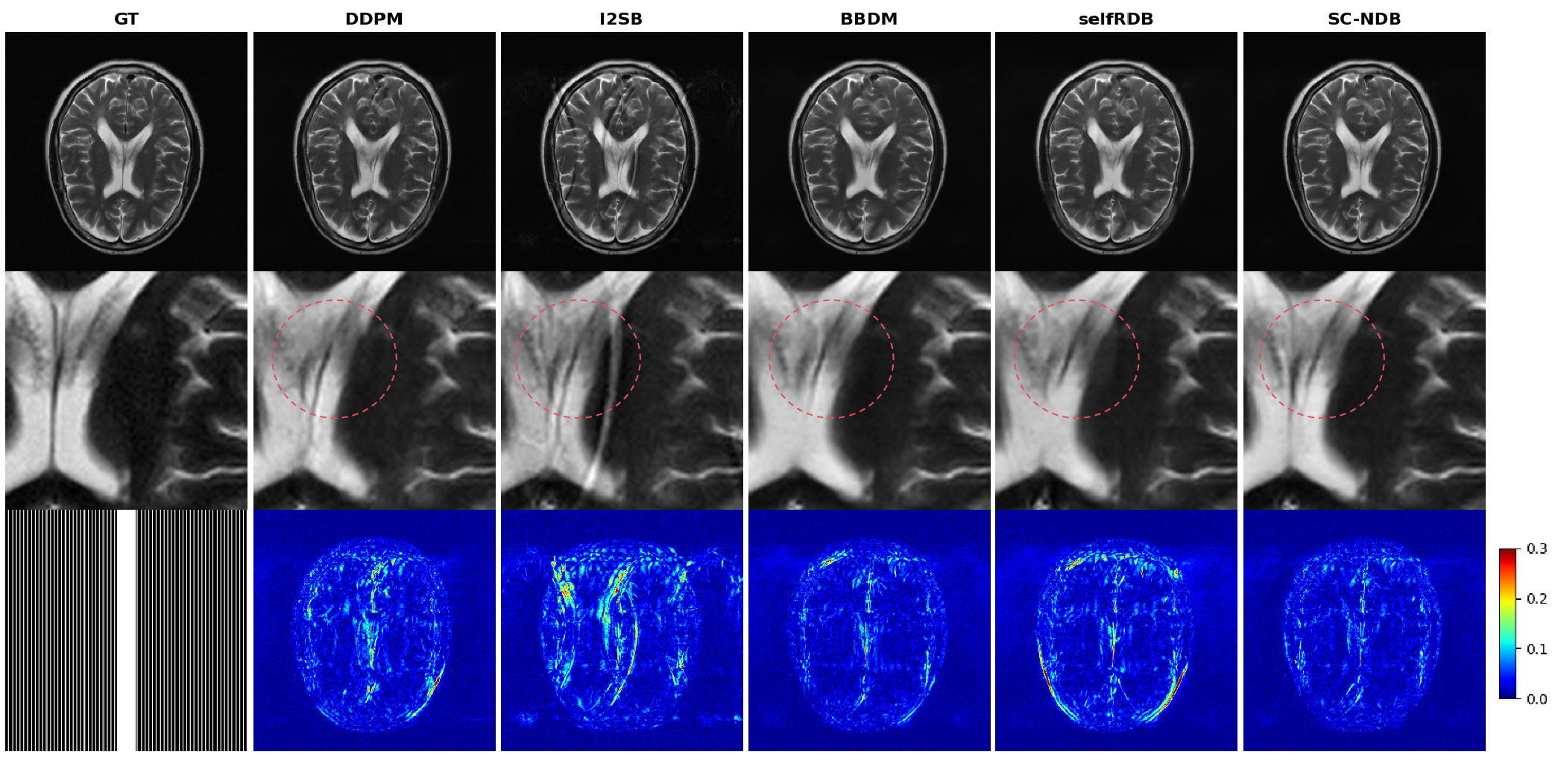}
    \caption{The reconstruction results of magnitude images generated from multi-coil brain fast MRI with a 4× acceleration factor under uniform Cartesian undersampling are presented (out-of-distribution). The first row displays the ground truth and the reconstructions from DDPM, I2SB, BBDM, selfRDB and SC-NDB (Ours). The second row presents the enlarged view of the ROI, and the third row shows the error map of the reconstruction. The lower left corner contains the undersampling mask.}
    \label{fig:brain_vis}
\end{figure*}

\begin{table}

\resizebox*{\columnwidth}{!}{%
\begin{tabular}{c|c|cccc}
\hline
Sampling    &Type                  & Method      & PSNR(dB)$\uparrow$   & SSIM(\%)$\uparrow$   & NMSE(\%)$\downarrow$  \\ \hline
 &    & AdaDiff\citep{gungor2023adaptive}         & 28.16±1.94 & 78.77±5.03 & 6.60±3.56 \\
                    &Non-magnitude    & MC-DDPM\citep{xie2022measurement}         & 33.84±3.38 & 88.24±3.57 & 3.02±3.05 \\
               &    & HFS-SDE\citep{high-frequency}         & 33.27±3.55 & 87.95±5.01 & 3.27±3.35 \\ \cline{2-6}
                      &  & DDPM\citep{DDPM}                 & 32.78±2.87           & 86.84±8.06           & 3.36±4.54            \\
                                $4\times$    &  & I2SB\citep{I2SB}                 & 33.45±2.54           & 87.49±6.53           & 2.59±3.08            \\
                                    &Magnitude  & BBDM\citep{bbdm}                 & 33.79±2.48           & 87.85±7.19           & 2.47±3.04            \\
                                    &  & selfRDB\citep{selfRDB}              & \underline{34.20±2.50}           & \underline{89.19±6.28}           & \underline{2.26±2.80}            \\
                        
                    &    & SC-NDB(Ours)        & \textbf{34.84±2.48} & \textbf{89.73±6.20} & \textbf{1.96±2.40} \\ \hline
  &  & AdaDiff\citep{gungor2023adaptive}              & 28.85±2.15           & 77.53±6.15           & 6.04±4.58            \\
                                    & Non-magnitude & MC-DDPM\citep{xie2022measurement}              & 31.30±2.45           & 84.58±5.84           & \underline{3.30±2.79}            \\
                                    &  & HFS-SDE\citep{high-frequency}              & 30.88±2.47           & 83.87±5.92           & 3.60±5.40            \\ \cline{2-6}
                                    &  & DDPM\citep{DDPM}                 & 30.36±2.79           & 81.45±9.06           & 5.44±8.59            \\
                                   $8\times$ &  & I2SB\citep{I2SB}                 & 31.45±2.25           & 82.54±7.73           & 3.78±4.03            \\
                                    &Magnitude  & BBDM\citep{bbdm}                 & 31.68±2.14           & 83.39±8.02           & 3.63±4.00            \\
                                    &  & selfRDB\citep{selfRDB}              & \underline{31.98±2.18}           & \underline{84.88±7.17}           & 3.41±3.76            \\

                    &    & SC-NDB(Ours)        & \textbf{32.80±2.20} & 
                        \textbf{85.83±6.97} & \textbf{2.96±3.37} \\ \hline
\end{tabular}%
}
\caption{Comparison of methods in in-distribution fastMRI knee experiments. $4\times$ and $8\times$ represent equally-spaced Cartesian undersampling with acceleration factors of 4 and 8, respectively. \textbf{Bold} and \textbf{underline} indicate the best and second-best results, respectively.}
\label{tab:knee}
\end{table}

\subsection{In-Distribution Results} 

\begin{figure}[tbp]
    \centering
    \includegraphics[width=1.0\textwidth]{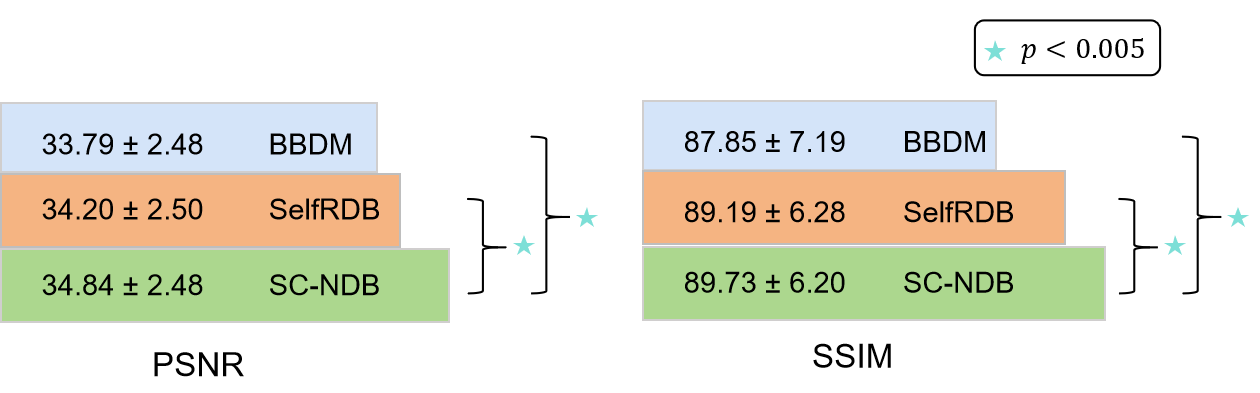}
    \caption{The Wilcoxon signed-rank test with a significance level of $p < 0.005$ was conducted to assess the statistical differences between SC-NDB and both BBDM and SelfRDB.}
    \label{fig:p}
\end{figure}

As presented in Table.~\ref{tab:knee} and Table.~\ref{tab:IXI}, our proposed SC-NDB demonstrates consistently superior performance on both the fastMRI Knee and IXI datasets under equally spaced Cartesian undersampling with acceleration factors of $4\times$ and $8\times$. Across all evaluation metrics, SC-NDB outperforms existing state-of-the-art methods on these in-distribution datasets. Specifically, compared to diffusion models based on magnitude images such as selfRDB~\cite{selfRDB} and BBDM~\cite{bbdm}, SC-NDB achieves an improvement of approximately 0.6–1.0 dB in PSNR and a 1–2$\%$ increase in SSIM. Furthermore, it maintains competitive performance when compared with approaches based on non-magnitude data~\cite{gungor2023adaptive, xie2022measurement, high-frequency}. As shown in Fig.~\ref{fig:p}, statistical significance tests on PSNR and SSIM metrics were conducted between SC-NDB and the BBDM and SelfRDB methods. The results indicate that SC-NDB demonstrates statistically significant improvements over both BBDM and SelfRDB, thereby validating the effectiveness of our proposed approach.

Fig.~\ref{fig:knee_vis} illustrates representative reconstruction results from the fastMRI knee dataset under $4\times$ equally spaced Cartesian sampling. The first row displays outputs from DDPM, I2SB, BBDM, selfRDB, and SC-NDB. The second row presents enlarged views of selected regions of interest (ROI), while the third row shows the corresponding error heatmaps. Reconstructions generated by DDPM, I2SB, BBDM, and selfRDB exhibit noticeable artifacts and noise, often leading to the loss of fine anatomical structures. In contrast, SC-NDB demonstrates superior reconstruction fidelity by preserving realistic high-frequency details and effectively suppressing noise and artifacts.

\begin{table}

\resizebox*{\columnwidth}{!}{%
\begin{tabular}{c|c|cccc}
\hline
Sampling    &Type                  & Method      & PSNR(dB)$\uparrow$   & SSIM(\%)$\uparrow$   & NMSE(\%)$\downarrow$  \\ \hline
 &    & AdaDiff\citep{gungor2023adaptive}         & 25.84±1.28           & 80.09±3.71           & 3.92±1.41  \\
                    &Non-magnitude    & MC-DDPM\citep{xie2022measurement}        & 36.20±3.11           & 97.49±1.31           & 0.37±0.30 \\
               &    & HFS-SDE\citep{high-frequency}         & 35.14±3.31           & 96.93±1.58           & 0.53±0.41 \\ \cline{2-6}
                      &  & DDPM\citep{DDPM}                 & 32.70±2.77           & 95.32±2.10           & 0.85±0.59           \\
                                $4\times$    &  & I2SB\citep{I2SB}                 & 34.31±2.76           & 96.23±1.71           & 0.58±0.41            \\
                                    &Magnitude  & BBDM\citep{bbdm}                & \underline{36.57±3.10}           & \underline{97.70±1.25}           & \underline{0.36±0.29}             \\
                                    &  & selfRDB\citep{selfRDB}              &  36.02±3.02           & 97.49±1.31           & 0.41±0.31            \\
                    &    & SC-NDB(Ours)        & \textbf{37.56±3.07} & \textbf{98.10±1.02} & \textbf{0.29±0.22} \\ \hline
  &  & AdaDiff\citep{gungor2023adaptive}               & 23.74±1.05           & 75.45±3.87           & 6.45±1.86            \\
                                    & Non-magnitude & MC-DDPM\citep{xie2022measurement}              & 35.53±2.96           & 97.04±1.53           & 0.45±0.34          \\
                                    &  & HFS-SDE\citep{high-frequency}              & 34.99±3.04           & 96.58±1.77           & 0.52±0.41            \\ \cline{2-6}
                                    &  & DDPM\citep{DDPM}                & 32.56±2.72           & 95.02±2.21           & 0.87±0.60            \\
                                   $8\times$ &  & I2SB\citep{I2SB}                 & 33.11±2.30           & 95.30±1.74           & 0.73±0.41            \\
                                    &Magnitude  & BBDM\citep{bbdm}                 & \underline{35.75±2.97}           & \underline{97.10±1.46}           & \underline{0.43±0.33}           \\
                                    &  & selfRDB\citep{selfRDB}              & 35.38±2.93           & 97.01±1.49           & 0.47±0.35           \\

                    &    & SC-NDB(Ours)        & \textbf{37.09±2.96} & 
                        \textbf{97.70±1.15} & \textbf{0.32±0.23} \\ \hline
\end{tabular}%
}
\caption{Comparison of methods in in-distribution IXI experiments. $4\times$ and $8\times$ represent equally-spaced Cartesian undersampling with acceleration factors of 4 and 8, respectively. \textbf{Bold} and \textbf{underline} indicate the best and second-best results, respectively.}
\label{tab:IXI}
\end{table}

\begin{table}

\resizebox*{\columnwidth}{!}{%
\begin{tabular}{c|c|cccc}
\hline
Sampling    &Type                  & Method      & PSNR(dB)$\uparrow$   & SSIM(\%)$\uparrow$   & NMSE(\%)$\downarrow$  \\ \hline
 &  & AdaDiff\citep{gungor2023adaptive}              & 24.93±1.30           & 81.21±3.07           & 13.2±4.37            \\
                                    &Non-magnitude  & MC-DDPM\citep{xie2022measurement}              & 29.93±1.76           & 90.47±2.14           & \underline{3.11±0.85}            \\
                                    &  & HFS-SDE\citep{high-frequency}              & 29.88±1.79           & 90.33±2.17           & 3.28±1.25            \\ \cline{2-6}
                    &  & DDPM\citep{DDPM}                 & 29.71±1.85           & 90.38±2.11           & 3.76±1.43            \\
                                 $4\times$   &  & I2SB\citep{I2SB}                 & 28.12±1.97           & 86.98±3.41           & 4.83±2.10            \\
                                    &Magnitude  & BBDM\citep{bbdm}                 & 29.41±1.64           & \underline{91.01±1.78}           & 3.16±1.12            \\
                                    &  & selfRDB\citep{selfRDB}              & \underline{29.78±1.64}           & 90.95±2.08           & 3.49±1.25            \\
                        
                    &    & SC-NDB(Ours)        & \textbf{30.67±1.71} & \textbf{91.97±1.42} & \textbf{2.98±0.94} \\ \hline
 &  & AdaDiff\citep{gungor2023adaptive}              & 25.67±1.37           & 80.08±3.13           & 10.0±2.72            \\
                                    &Non-magnitude  & MC-DDPM\citep{xie2022measurement}              & 26.10±2.21           & 82.52±3.56           & 7.91±2.89            \\
                                    &  & HFS-SDE\citep{high-frequency}              & 26.36±2.10           & 83.35±2.32           & 7.35±1.62            \\ \cline{2-6}
                &  & DDPM\citep{DDPM}                 & 27.47±1.30           & 86.93±2.23           & 6.15±1.56            \\
                                   $8\times$ &  & I2SB\citep{I2SB}                 & 25.77±1.30           & 82.48±4.00           & 7.67±2.16            \\
                                    &Magnitude  & BBDM\citep{bbdm}                 & \underline{27.56±1.30}           & 86.95±2.19           & 6.35±1.64            \\
                                    &  & selfRDB\citep{selfRDB}              & 27.46±1.24           & \underline{87.46±2.66}           & \underline{6.04±1.71}            \\

                    &    & SC-NDB(Ours)        & \textbf{28.65±1.24} & 
                        \textbf{89.02±1.94} & \textbf{4.62±1.20} \\ \hline
\end{tabular}%
}
\caption{Comparison of methods in out-of-distribution fastMRI brain experiments, using models trained on the in-distribution fastMRI knee dataset. $4\times$ and $8\times$ represent equally-spaced Cartesian under-sampling with acceleration factors of 4 and 8, respectively. \textbf{Bold} and \textbf{underline} indicate the best and second-best results, respectively.}
\label{tab:brain}
\end{table}

\begin{table}[]
\centering
\begin{tabular}{llccc}
\hline
Self-Consistency      & CDEM                  & PSNR(dB)$\uparrow$   & SSIM(\%)$\uparrow$   & NMSE(\%)$\downarrow$  \\ \hline
                      &                       & 33.79±2.48 & 87.85±7.19 & 2.47±3.04 \\
                      & \multicolumn{1}{c}{\checkmark} & 34.15±2.53 & 88.25±7.19 & 2.32±2.86 \\
\multicolumn{1}{c}{\checkmark} &                       & 34.62±2.49 & 89.56±6.19 & 2.04±2.47 \\
\multicolumn{1}{c}{\checkmark} & \multicolumn{1}{c}{\checkmark} & \textbf{34.84±2.54} & \textbf{89.73±6.20} & \textbf{1.96±2.40} \\ \hline
\end{tabular}
\caption{Ablation studies of different components. We report the metrics on the fastMRI knee dataset at a 4× acceleration factor.}
\label{ablation:compoents}
\end{table}

\begin{table}[]
\centering
\begin{tabular}{llccc}
\hline
$t1=t2\sim U$                  & $t1\sim U,t2\sim U$                 & PSNR(dB)$\uparrow$   & SSIM(\%)$\uparrow$   & NMSE(\%)$\downarrow$  \\ \hline
\multicolumn{1}{c}{\checkmark} &                       & 34.84±2.54 & 89.73±6.20 & 1.96±2.40 \\
                      & \multicolumn{1}{c}{\checkmark} & 34.85±2.55 & 89.75±6.16 & 1.96±2.39 \\ \hline
\end{tabular}
\caption{Ablation studies of different time steps in $\boldsymbol{x}_0/\boldsymbol{y}_0\to \boldsymbol{y}_0/\boldsymbol{x}_0$ and $\boldsymbol{y}_0/\boldsymbol{x}_0\to \boldsymbol{\bar{x}_0}/\boldsymbol{\bar{y}_0}$. $U$ denotes a uniform distribution $Uniform(1,...,T)$.}
\label{ablation:time}
\end{table}

\begin{table}[]
\centering
\begin{tabular}{lccc}
\hline
$\sqrt{\sigma_{t2}}$                 & PSNR(dB)$\uparrow$   & SSIM(\%)$\uparrow$   & NMSE(\%)$\downarrow$  \\ \hline
   1.0                     & 34.67±2.50	&89.54±6.20	&2.03±2.49 \\
                      1.1 & 34.79±2.54 & 89.73±6.19 & 1.97±2.41 \\ 
 1.2 & \textbf{34.84±2.54} & \textbf{89.73±6.20} & \textbf{1.96±2.40} \\ 

    1.3 & 34.76±2.56	&89.70±6.19	&2.00±2.47 \\ \hline

\end{tabular}
\caption{Ablation study was performed by varying the noise variance using $\sqrt{\sigma_{t2}}$ to evaluate its impact on reconstruction performance.}
\label{ablation:var}
\end{table}

\subsection{Out-of-Distribution Results}
We further conducted out-of-distribution experiments to verify the generalization of our model. All models were trained on fastMRI knee data and tested by fastMRI brain MRI scans. The reconstruction results for equally-spaced Cartesian sampling with $4\times$ acceleration are shown in Fig.~\ref{fig:brain_vis}. The reconstruction quality of DDPM, I2SB, BBDM and selfRDB methods significantly decreased, whereas diffusion-based methods maintained good reconstruction quality. Among them, our SC-NDB achieved the optimal reconstruction results with fewer noise and artifacts. Table~\ref{tab:brain} gives quantitative performance for the out-of-distribution experiments and our proposed SC-NDB model achieves the best performance for the brain data reconstruction.
\subsection{Inference Efficiency}
\begin{figure}[tbp]
    \centering
    \includegraphics[width=1.0\textwidth]{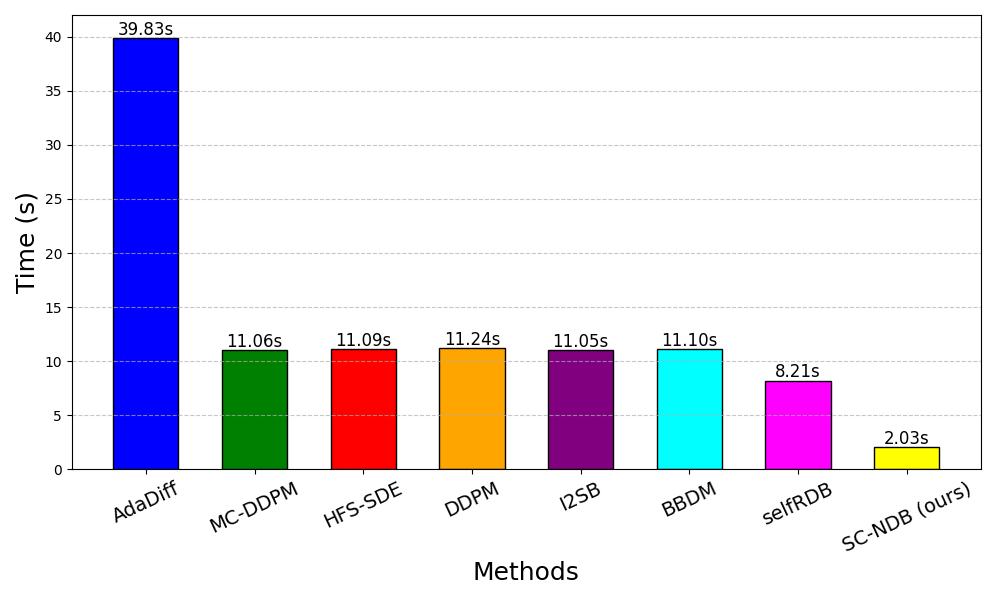}
    \caption{A bar chart comparing the inference time of various diffusion methods.}
    \label{fig:infer_time}
\end{figure}

In the above experiments, the training and sampling steps for SC-NDB and selfRDB were set to 20 each, while other diffusion methods (except AdaDiff) used 1000 training steps and 200 sampling steps. The 200 sampling steps were accelerated using DDIM~\cite{ddim} sampling. For AdaDiff, the first stage involved pretraining with 1000 steps, and the second stage did not use 200 sampling steps but instead performed 200 online iterations based on the data consistency loss. As shown in Fig.~\ref{fig:infer_time}, SC-NDB's significantly reduced sampling steps result in much higher inference efficiency compared to other diffusion methods.

\subsection{Ablation Study}
Table~\ref{ablation:compoents} shows an ablation study to verify the effectiveness of our designs. The traditional BBDM \cite{bbdm} model is set as the baseline. Results indicate that BBDM achieves better results than other diffusion models. Then, both the self-consistency and CDEM designs improve the performance of the knee MRI reconstruction. Meanwhile, higher SSIM results indicate better structural fidelity of the reconstruction results. Table~\ref{ablation:time} presents a comparative analysis of the settings for $t1$ and $t2$ during the training process of our SC-NDB. Similarly, Table~\ref{ablation:var} compares the effects of different settings for $\sqrt{\sigma_{t2}}$ under $\sqrt{\sigma_{t1}}=1$. The results reveal that the time steps and noise variances for the two nested diffusion bridge processes can vary in each step of the diffusion bridge process, highlighting the versatility and adaptability of the SC-NDB training approach.
\subsection{Limitations}
Although our SC-NDB framework enhances the modeling of source image distributions, the training procedure still necessitates the simultaneous optimization of two nested diffusion bridges, potentially leading to increased computational costs. Moreover, as shown in Table~\ref{tab:brain}, the performance degrades under out-of-distribution scenarios. Future work will focus on developing more computationally efficient diffusion models with improved generalization capabilities.

\section{Conclusion}
In this study, we proposed the Self-Consistent Nested Diffusion Bridge (SC-NDB), a novel diffusion-based framework tailored for magnitude-image-based accelerated MRI reconstruction. Unlike conventional approaches relying on complex-valued image-space or k-space data, SC-NDB focuses on the more clinically accessible magnitude images, which are widely adopted in real-world applications. By introducing a nested structure of bidirectional bridge diffusion processes and enforcing self-consistency during training, SC-NDB more effectively captures explicit priors from the source image, leading to improved translation accuracy between under-sampled and fully-sampled MRI data.

Additionally, the proposed Contour Decomposition Embedding Module (CDEM) enhances the model’s ability to retain fine anatomical structures by leveraging multi-scale and directional structural information. Extensive experiments on the fastMRI and IXI datasets demonstrate that SC-NDB consistently outperforms both state-of-the-art magnitude-based and non-magnitude-based diffusion models in terms of reconstruction quality and fidelity.

Our findings highlight the potential of SC-NDB as a practical and effective solution for real-world MRI reconstruction. Future work will explore further optimization of computational efficiency and generalization capabilities, particularly in out-of-distribution scenarios and broader clinical settings.



\bibliographystyle{model2-names.bst}\biboptions{authoryear}
\bibliography{refs}
\end{document}